\documentclass[12pt]{article}
\usepackage[margin=1in]{geometry}
\usepackage{graphicx}
\newcommand{\vv}{\vspace*{1.5ex}}

                            \newcommand{\no}{\noindent}
 \newcommand{\bc}{\begin{center}}
 \newcommand{\ec}{\end{center}}
                   \newcommand{\bfr}{\begin{flushright}}
                   \newcommand{\efr}{\end{flushright}}
   \newcommand{\ii}{\item}
     \newcommand{\be}{\begin{enumerate}}
     \newcommand{\ee}{\end{enumerate}}
        \newcommand{\bi}{\begin{itemize}}
        \newcommand{\ei}{\end{itemize}}
            \newcommand{\bd}{\begin{description}}
            \newcommand{\ed}{\end{description}}
                \newcommand{\beq}{\begin{equation}}
                \newcommand{\eeq}{\end{equation}}
                  \newcommand{\bea}{\begin{eqnarray}}
                  \newcommand{\eea}{\end{eqnarray}}

      \newcommand{\bfi}{\begin{figure}}
      \newcommand{\efi}{\end{figure}}
\newcommand{\bay}{\begin{array}{l}}
\newcommand{\eay}{\end{array}}
            \newcommand{\dd}{\mbox{d}}

    \newcommand{\del}{\delta}
    \newcommand{\Del}{\Delta}

    \newcommand{\al}{\alpha}

    \newcommand{\tht}{\theta}  
    \newcommand{\ga}{\gamma}

\begin{document} 
\thispagestyle{empty}
        \hspace*{1mm}  \vspace*{-0mm}
\noindent {\footnotesize {{\em
\hfill      Posted on ArXiv under identifier: submit/2522317} }}
    \vskip 1.5in
\begin{center}
{\large {\bf     Direct Multilayer Adsorption of Vapor in Solids with 
\\[1.7mm]  Multiscale Porosity and Hindered Adsorbed Layers in Nanopores 
            }}\\[15mm]

{\large {\sc Zden\v ek P. Ba\v zant and Hoang Thai Nguyen}}
\\[1.2in]

{\sf SPREE Report No. 18-12/33788d}\\[2.1in]

Center for Science and Protection of Engineering Environment (SPREE)
\\ Department of Civil and Environmental Engineering
\\ Northwestern University
\\ Evanston, Illinois 60208, USA
\\[1in]  {\bf December 27, 2018} 
\end{center}

\clearpage   \pagestyle{plain} \setcounter{page}{1}
  \vskip 1.5in


\begin{center}
                     {\large {\bf {\sf
           Direct Multilayer Adsorption of Vapor in Solids with 
\\[1.7mm]  Multiscale Porosity and Hindered Adsorbed Layers in Nanopores 
 }}} \\[7mm]

{\large {\sf
    Zden\v ek P. Ba\v zant}\footnote{
McCormick Institute Professor and W.P. Murphy Professor of Civil
and Mechanical Engineering and Materials Science, Northwestern University, 2145
Sheridan Road, CEE/A135, Evanston, Illinois 60208; corresponding author,
z-bazant@northwestern.edu.}  and {\sf
    Hoang Thai Nguyen}\footnote{Graduate Research Assistant, Northwestern University}}
\end{center} \vskip 3mm 

\noindent {\bf Abstract:}\, {\sf Hindered adsorbed layers completely filling the nanopores must cause significant deviations from the classical BET isotherms for multimolecular adsorption of vapor in porous solids. Since the point of transition from free to hindered adsorption moves into wider nanopores 
adsorption layer exposed to vapor gets reduced by an area reduction factor that decreases with increasing adsorbed volume, and thus also with increasing vapor pressure (or humidity). The area reduction factor does not affect the rates of direct adsorption or condensation from individual vapor/gas molecule, which represent a local process, but imposes a lateral constraint on the total area and volume of the free portion of the adsorption layer that is in direct contact with vapor. Assuming an inverse power law for the dependence of the area reduction factor on the number of molecular layers, one can express the modified isotherm in terms of logarithmic or polylogarithm (aka Jonqui\`ere) functions. The power-law exponent is a property that serves as an additional data fitting parameter. For the same initial slope, the modified isotherms deviate from the BET isotherm downwards, and the deviation increases with the exponent. Comparisons with some published isotherms obtained experimentally on cement pastes show that the present modification of the BET theory for hindered adsorption goes in the right direction. Detailed calibration of the theory and an extension for indirect communication of vapor molecules with the molecules adsorbed in nanopores less than a few nm wide will require further research.   }

\vv \no Key Words: Adsorption isotherm, free adsorption, BET theory, hindered adsorption, evaporation and condensation, variation of adsorption surface, statistical analysis, capillary condensation theory, pore size distribution, polylogarithm, Jonqui\`ere functions.  

\subsection*{Introduction and Basic Concepts}

Adsorption of gases or vapors in multimolecular layers in porous solid is generally described by the classical BET theory, formulated in 1938 by Brunauer, Emmett and Teller (rumored to stem from a back-of-the envelope calculation of Edward Teller during a lunch) \cite{brunauer1938adsorption, brunauer1940theory, brunauer1943adsorption}.
One important application of the BET has been the water desorption and 
adsorption 
in cement hydrates and concrete, which is here the main application in mind, although a similar problem arises, e.g., in activated carbon fibers formed by crystallite graphite sheets \cite{kaneko1992origin}. 
Various useful improvement of the BET theory, particularly its extensions to the capillary range, have been formulated \cite{lykov1958transporterscheinungen,freiesleben1985coupled,kunzel1995simultaneous,brunauer1940theory,brunauer1969adsorption,halsey1948physical,hillerborg1985modified,xi1994moisture,bavzant2018creep}, but the effect of varying nanopore sizes on the area exposed to vapor, important for water in concrete, seems to have eluded attention.

The BET theory assumes that the number of molecular layers is unhindered, which requires wide enough pores. The number of adsorbed water layers containing significant mass increases with the vapor pressure, $p$, or relative humidity $h = p /p_{sat}$ and, for high enough $h$, there can be up to 5 layers with significant adsorbed mass, which gives the approximate maximum thickness of 1.335 nm ($p_{sat} = p_{sat}(T)$ = saturation vapor pressure, a function of $T$). This means that, in nanopores less than 2.67 nm wide (which constitute the major part of pore volume in cement hydrates), and for not too small $h$, the adsorption layers on the opposite pore surfaces touch and fill the pore completely. 

In materials with multiscale nanoporosity, such as concrete, the maximum thickness of nanopores filled by adsorbed water molecules decreases with decreasing $h$ (Fig. \ref{fig1}b,c). When $h$ exceeds the value at the limit of complete filling, the adsorption layer is hindered from developing its full thickness, and such a hindered adsorbed water layer (sometimes less fittingly also called the "interlayer water") develops a significant transverse pressure, called the disjoining pressure \cite{deryagin1940theory, powers1966some, bavzant1972thermodynamics, bavzant1972nuclear,bavzant2012theory-1, bavzant2012theory-2}. 
The hindered adsorbed water requires us to distinguish two simple limiting cases:
 \be \setlength{\itemsep}{-1.6mm} 
 \ii
  In thin enough nanopores, less than a few nanometers thick, the adsorbed water molecules are restricted by solid surface forces and cannot communicate with the vapor in larger (capillary) pores directly. 
  The adsorbed water molecules can communicate with water vapor in the larger pores only indirectly, by surface diffusion along the adsorption layers. The surface diffusion process is far slower than the process of direct adsorption from vapor on the adsorption layer surface exposed to vapor. Consequently, for the wetting or drying of porous specimens, which are sufficiently thin so that the delay due to macro-diffusion through the specimen is negligible, the water in the nanopores, whether filling the nanopores fully or partially, may be considered to be almost immobile. In this limiting case, which will be the object of the present analysis, the mathematical generalization of the BET theory is not difficult. One can ignore diffusion along the solid surface of the nanopores and deal only with the direct exchange of the adsorbate (water) molecules between the vapor and adsorption layers (which is why the term "direct" appears in the title).
 \ii
  The opposite case is a very slow drying or wetting, which allows the surface diffusion of hindered adsorbed water along the nanopores to run its course and come to a standstill. This limiting case should also be analytically tractable, in a way suggested in the Appendix. But more research is needed.
 \ee


The mathematical derivation of the BET theory \cite{brunauer1938adsorption,brunauer1943adsorption} is valid only for free, rather than hindered adsorbed layers
  \cite{bavzant2018creep}.  
This derivation is valid only when the surface, of area $A'$, of the multimolecular adsorption layer in contact with the vapor is equal to be adsorbent area $A$ and is independent of the number of layers. However, in materials such as cement hydrates, there are nanopores of highly variable width (Fig. \ref{fig1}b) and nanopores of uniform but very different widths (Fig. \ref{fig1}c). Consequently, the vapor exposed area, $A'$, in cement hydrates and some other nanoporous materials must decrease significantly as $h$ increases. In the limit case of sufficiently fast wetting or drying, in which the filling of the thin nanopores cannot change significantly, as stipulated at the outset (case 1), the decrease of $A'$ may be described by an {\em area reduction factor}, $\beta_n$, which depends on the number, $n$, of multimolecular layers and reduces the full area $A$ of the bare (or dry) internal pore surface per unit volume of the porous material (Fig. \ref{fig1}a), i.e.
 \beq \label{a1}
         \fbox{$\displaystyle ~
               A'\ =\ \beta_n\, A         ~ $}
                          ~~~~(n = 0, 1,2,3,...,~~ \beta_0 = 1)
 \eeq
Here $\beta_n$ is a decreasing sequence; see Fig. \ref{fig1} (in cement paste, $A \approx$ 500 m/cm$^3$, which implies the average pore width to be about $0.5$ nm \cite{le2011unified}, or about two water molecules).

As for the hindered adsorption layers, which fill the nanopores completely (and are not thicker than 2.67 nm), we may assume that the migration of adsorbate (water) molecules along these layers (Fig. \ref{fig1}b,c) is so slow that it does not intervene appreciably with the rates of adsorption, evaporation and condensation on the surface in contact with vapor.

The decrease of the surface area of multimolecular adsorption layers with
increasing number $n$ of the layers may be schematically represented as shown in Fig. \ref{fig1}a. Each horizontal line represents adsorbate volume that is equivalent to one molecular layer (the regular arrangement of molecules is, of course, only a mean idealization of a constantly varying random arrangement of molecules). The first layer occupies area $A$ of the bare adsorbent surface. For increasing $n$, we imagine the dashed lateral constraint to reduce the area of each molecular layer from $A$ to $A' = \beta_n A$, where $\beta_n$ ($n = 1,2,3,...$) is a monotonically decreasing sequence, such that $\beta_0$ = 1 and $\beta_n > 0$ for all $n$.

A certain restriction on the surface area exposed to vapor was formulated, in 1940 by Brunauer {\em et al.}'s \cite{brunauer1940theory}. They generalized the BET theory to take into account two-sided  adsorption layers on opposite parallel planar walls of a nanopore. The interference of the opposite twp-sided adsorption layers causes that the area exposed to vapor in the nanopore decreases as the adsorbate volume increases and vanishes as the nanopore gets full. A decrease of the area exposed to vapor also features in the present formulation, but that is where the similarity ends. Here, most of the increase of the exposed area with the multilayer thickness has a different source, explained by Fig. \ref{fig2}b,c. In \cite{brunauer1940theory}. This increase is achieved not by growth of the adsorbate volume in filled nanopores of varying width, but by a denser random filling of a nanopore of constant width (a pore between two parallel plates; see Fig. \ref{fig2}b, or Fig. 4 in \cite{brunauer1940theory}. The main problem with \cite{brunauer1940theory} is the disregard of the vastly increased resistance to the movement of adsorbate molecules along the nanopore, which occurs by surface diffusion. Unlike here, the statistical analysis in \cite{brunauer1940theory} implies nanopores of uniform width, which is true for some materials (e.g., charcoal or crystalline dolomite rocks) but is far from true for the cement hydrates. As another difference, in \cite{brunauer1940theory} it is considered that when the molecular layers growing from opposite surfaces of a planar nanopore touch, their heat of liquefaction increases, while this increase is not modeled. The calculated isotherm, given by equations (E) and 16 in \cite{brunauer1940theory}, is much more complicated than what is obtained here. Brunauer {\em et al.}'s aim was to describe the isotherms of types IV and V, as defined in figure 1 of  \cite{brunauer1940theory, brunauer1943adsorption},
while here we aim at the isotherms of types I, II and III (although a simple adjustment could also fit types IV and V). The theory in \cite{brunauer1940theory}, based on statistics of adsorption in a planar nanopore of constant width, cannot be extended to the capillary range, while the present theory can.  


\subsection*{Adsorption under Lateral Constraint of Hindered Adsorption Layers}

Adsorption is a random process in which water molecules constantly enter the adsorption later, stay there for a certain time called the lingering time (about $10^{-9}$ s \cite{deboer1953dynamical}) and then exit into the vapor. In equilibrium, the number of water molecules within the layer is at any time is, macroscopically, exactly the same. Although the arrangement of molecules at any moment varies and looks rugged as shown in Fig. \ref{fig1}c, one can define a precise effective thickness $\del_{ef}$ as the volume occupied by all the water molecules per unit base area.

Let us now follow as closely as possible the original derivation of the BET theory \cite{brunauer1938adsorption, brunauer1943adsorption}, though with some vital differences. With a focus on hydrated cement, let us consider that the adsorbate is water, although it could be some other substances. Let $A$ be the total base area of the adsorbent base, corresponding to the internal pore surface in a porous materials, and let $\al_n$ be the exposed coverage density in the $n^{th}$ adsorption layer, i.e., the area fraction of the adsorbed (or water) molecules that are exposed to vapor in the $n^{th}$ layer (Fig. \ref{fig1}a). 

Consider now a small area $\dd A$ of the adsorbent base. The rate (or probability) of the water vapor molecules condensing (or getting adsorbed) on top of the $(n-1)^{st}$ layer is $q_1\, \dd A = a_n \al_{n-1} p\, \dd A$  where $p$ is the vapor pressure and $a_n$ is a constant (to be determined later).

The  rate (or probability) of the vapor molecules condensing from the vapor into the $n^{th}$ layer (i.e., on top of the $(n-1)^{th}$ layer), is $q_2\, \dd A = b_n \al_n e^{- Q/RT} \dd A$ where $b_n$ is another constant (to be determined later), $T$ = absolute temperature, $R$ = gas constant, and $Q$ = activation energy. For the first layer, $n=1$, $Q = Q_a$ = heat of adsorption (which is dissipated upon upon severance of van der Waals bonds at the adsorbent surface as water molecules escape from the surface). For all the layers except the first, it is assumed, same as in the BET theory, that $Q = Q_l$ = heat of liquefaction, which means that the van der Waals forces of the adsorbent surface are assumed to have no appreciable effect beyond the first layer.
 
In thermodynamic equilibrium, both rates (or probabilities) must be equal, i.e., $q_1 = q_2$, and so
 \bea  \label{a2}
   \mbox{for}~~n = 1:~~~~~~~~~~~~~~~a_1 \al_0 p &=& b_1 \al_1 e^{-Q_a /RT}
 \\  \label{a3}
   \mbox{for}~~n = 2,3,4,...:~~~a_n \al_{n-1} p &=& b_n \al_n e^{- Q_l/RT}~~~~ \eea
which means that, for each layer, the density (or probability) $q_1$ of the rate of adsorption or condensation must be equal to and the density (or probability) $q_2$ of the rate of evaporation. 

Note that the area reduction factors, $\beta_n$, must not appear in these equations because they represent densities, per unit area of the base layer, and are independent of the lateral spread of that layer defined by these factors. However, these area reductions factors, which represent lateral constraints on the layer areas (Fig. \ref{a3}), intervene in the global constraints, which are:
 \bea  \label{a4}
  && A = \sum_{n=0}^\infty \beta_n (\al_n A)
 \\  \label{a5}
  && v = \del_1 \sum_{n=0}^\infty n \beta_n (\al_n A)
 \eea 
where $\del_1$ = effective thickness of monomolecular adsorbed layer (representing the volume, $v_m$, of that layer per unit area of adsorbent surface), and $v$ = total volume of adsorbate over area $A$. Eq. (\ref{a4}) means that the sum or the top areas of the random columns of $n$ water molecules, for all possible column heights $n$ (Fig. \ref{fig2}c), must be equal to the adsorbent base area $A$.       
Eq. (\ref{a5}) sums the volumes of all the columns of water molecules of various random heights $n$ (the volume and area per water molecule are omitted in these equations because they later cancel out). The  ratio   
defines the effective thickness of the adsorption layer. Taking the dimensionless ratio $v / A \del_1$ and setting  $A \del_1 = v_m$ = volume of a full monomolecular layer, we obtain the overall lateral constraint:
 \beq \label{a6}     
  \frac v {A \del_1} = ~ \fbox{$\displaystyle ~ \frac v {v_m} = \frac{
  \sum_{n=0}^\infty n \beta_n \al_n}{\sum_{n=0}^\infty \beta_n\al_n}   ~$}
 \eeq
which represents the key difference from the BET theory \cite{brunauer1938adsorption} and its extension to two-sided adsorption \cite{brunauer1940theory}.

Assuming that the van der Waals forces of solid surface do not reach beyond the first molecular layer, and noting that a molecule entering of the layers beyond the first essentially undergo liquefaction, we may consider the ratios $b_n/\al_n$ to be constant for $n > 1$, i.e.,
 \beq \label{a7}
  \frac{b_2}{a_2} = \frac{b_3}{a_3} = \frac{b_2}{a_2} = ... = g
 \eeq
where $g$ is a constant. Let us now define two variables:
 \bea \label{a8}
  && y = \frac{a_1}{b_1}\ p\, e^{Q_a/RT}
 \\  \label{a9}
  && h = \frac p g \ e^{Q_l/RT}
 \eea
(as shown later, $h = p /p_{sat}$ = relative humidity of the vapor). With these
notations,
 \bea  \label{a10}
  && \al_1 = y \al_0
 \\ \label{a11}
  && \al_n = h \al_{n-1} = h^2 \al_{n-2} = h^3 \al_{n-3} = h^{n-1}\al_1
         = y h^{n-1} \al_0 = c_T h_n \al_0~~~~~~~~~~~~
 \\  \label{a12}
  && \mbox{in which}~~~~~~
  c_T = \frac y h = \frac{a_1}{b_1}\,g\, e^{\Del Q /RT},~~~~\Del Q = Q_a - Q_l
 \eea
where always $Q_a > Q_l$. The saturation humidity, $p = p_{sat}$, is obtained for $v \to \infty$.      
This corresponds to $h = 1$ and shows that the meaning of notation $h$ is indeed the relative humidity of pore vapor, i.e.,
 \beq \label{a13}
  h = \frac p {p_{sat}}
 \eeq
and also that, according to Eq. (\ref{a9}), $(p_{sat} / g) e^{Q_l /RT} = 1$ or
 \beq \label{a14}
  g = p_{sat}(T)\, e^{Q_l /RT}
 \eeq
So we may conclude that
 \beq \label{a15}
  c_T = c_0 \, e^{Q_l /RT}
 \eeq
where $c_0$ is an empirical calibration parameter (close to 1).

Substitution of Eq. (\ref{a10}) and (\ref{a11}) into Eq. (\ref{a6}) now furnishes for the sorption isotherm the result:
 \beq  \label{a16}
    \frac v {v_m} = ~        \fbox{$\displaystyle ~
  \tht(h,T) = \frac{ c_T \sum_{n=1}^\infty \beta_n\, n\, h^n}
                   { 1 + c_T \sum_{n=1}^\infty \beta_n\, h^n }        ~ $}
 \eeq
For the special case that $\beta_n$ = 1 for all $n$ (no hindered adsorption), this formula yields the BET isotherm.

\subsection*{Isotherms Obtained for Various Area Reduction Factors}

Because of their self-similarity, power functions of $n$ appear to be a suitable choice for $\beta_n$ and, as it turns out, the infinite sums can be evaluated analytically. The simplest choice, satisfying the condition that $\beta_0$ = 1, is  \beq \label{a17}
  \beta_n\, =\, \frac 1 {1 + n}
 \eeq
for which
 \bea \label{a18}
  && \sum_{n=1}^\infty \beta_n h^n = \sum_{n=1}^\infty \frac{h^n}{1+n}
  = -\ \frac{h+\ln(1 - h)} h
 \\ \label{a19}  \mbox{and}~~
  && \sum_{n=1}^\infty \beta_n n h^n = \sum_{n=1}^\infty \frac{n h^n}{1+n}
  = -\frac{h\ln(1-h)-h-\ln(1-h)}{h(1-h)}~~~~~~~~~
 \eea
 Eq. (\ref{a16}) then yields, for the hindered isotherm, the final expression (Fig. \ref{fig3}a):
 \beq \label{a20}
  \tht(h,T)\ =\ c_T\,\frac{ h + \ln H-h\ln(H) }{H (h-c_T \ln H - c_T h)}\, ,~~~~H = 1-h
 \eeq

Fig. \ref{fig3}a shows a comparison of the plot of this hindered isotherm with the BET isotherm. Note that, similar to the BET isotherm, this isotherm cannot be valid in the capillary range ($h > 0.85$),  
unless a finite number $n$ is considered. This is obvious since $\lim_{h \to 1}~\tht = \infty$.

For the $5^{th}$ molecular layer, the area reduction factor due to hindered adsorption is $\beta_5$ = 1 /16. Whether this is realistic will depend on experiment (or MD simulations).

Eq. (\ref{a17}) for $\beta_n$ does not have any fitting parameter to optimize the fit of experimental adsorption data, and so it would be by luck if it provided a good fit of some experimental data. More generally, we can introduce a fitting parameter, $r$, such that
 \beq \label{a21}
  \beta_n\, =\, \frac 1 { (1 + n)^r }
 \eeq
Varying $r$, one has a continuous transition to the BET theory, which is attained for $r \to 0$. For arbitrary $r$, the isotherm can be expressed in terms of a special function, the polylogarithm (aka Jonqui\`ere's function), which is denoted by $Li_{r}(h)$ (and is valid when $|h|<1$). We have (Fig. \ref{fig3}a):
\bea \label{a22}
  && \sum_{n=1}^\infty \beta_n h^n = \sum_{n=1}^\infty \frac {h^n}{(1+n)^r} = \sum_{n=0}^\infty \frac {h^n}{(1+n)^r}-1
  = \frac{Li_r(h)}{h} - 1
 \\ \label{a23}  \mbox{and}~~
  &&\sum_{n=1}^\infty \beta_n n h^n = \sum_{n=1}^\infty \frac{n h^n}{(1+n)^r}
  = \frac{Li_{r-1}(h)-Li_{r}(h)}{h}~~~~~~~~~
 \\ \label{a24}  \mbox{Hence}~~~&&~~~~~ 
 \tht(h,T)\ =\ c_T\,\frac{Li_{r-1}(h)-Li_{r}(h)}{h+c_T(Li_{r}(h)-h)}
 \eea

Another, simpler, way is to assume $\beta_1$ ($\le 1$) to be a free fitting parameter, and consider that the subsequent area reduction factors are:
 \beq \label{a25}
  \beta_n = \frac{\beta_1}{n}, ~~~~ n = 1,2,3,4,...
 \eeq
As a result, the summed series and $\tht$ function are followed (Fig. \ref{fig3}b):
\bea \label{a26}
  && \sum_{n=1}^\infty \beta_n h^n = \sum_{n=1}^\infty \frac {\beta_1 h^n}{n} = -\beta_1 \ln(1-h)
 \\ \label{a27}  
  && \sum_{n=1}^\infty \beta_n n h^n = \sum_{n=1}^\infty \frac {\beta_1 n h^n}{n} = \beta_1 \frac{h}{1-h}~~~~~~~~~
  \\ \label{a28}  \mbox{Hence}~~~&&~~~~~ 
 \tht(h,T)\ =\ \frac{c_T \beta_1 h}{(1-h)\left(1-c_T \beta_1\ln(1-h) \right)}
 \eea

Another alternative is to add another fitting parameters $r$ together with $\beta_{1}$, which would be:
 \beq \label{a29}
  \beta_n = \frac{\beta_1}{ n^r }, ~~~~ n = 1,2,3,4,...
 \eeq
For $\beta_1 \to \beta_0 = 1$ and $r \to 0$, one can have again a continuous transition to the BET theory (Fig. \ref{fig3}b).
\bea \label{a30}
  && \sum_{n=1}^\infty \beta_n h^n = \sum_{n=1}^\infty \frac {\beta_1 h^n}{n^r} = \beta_1 Li_r(h)
 \\ \label{a31}  \mbox{and}~~
  && \sum_{n=1}^\infty \beta_n n h^n = \sum_{n=1}^\infty \frac {\beta_1 n h^n}{n^r} = \beta_1 Li_{r-1}(h)~~~~~~~~~
  \\ \label{a32}  \mbox{Leading to}~~&&~~ 
 \tht(h,T)\ =\ \frac{c_T \beta_1 Li_{r-1}(h)}{1+c_T \beta_1 Li_r(h) }
 \eea

If the experimental (or MD) data are abundant enough to optimize more fitting parameters, $\beta_1$, $\beta_2$, and $r$, one may consider that
\beq \label{a33}
  \beta_n = \frac{\beta_2}{ (n-1)^r }, ~~~~ n = 2,3,4,5,...
\eeq
For $r = 1$:
\bea \label{a34}
  && \sum_{n=1}^\infty \beta_n h^n = \beta_1 h + \sum_{n=2}^\infty \frac {\beta_2 h^n}{n-1} = \beta_1 h - \beta_2 h \ln(1-h)
 \\ \label{a35}  \mbox{and}~~
  && \sum_{n=1}^\infty \beta_n n h^n = \beta_1 h + \sum_{n=2}^\infty \frac {\beta_2 nh^n}{n-1}
  = \beta_1 h + \beta_2\frac{h+h\ln(1-h)-\ln(1-h)}{1-h}~~~~~~~~~
 \eea
Therefore, the sorption isotherm can be obtained as (Fig. \ref{fig3}c):
\bea \label{a36}
  \tht(h,T)\ =\ c_T\,\frac{ \beta_1 h H + \beta_2(h+h\ln H -\ln H )}{(1-h)[1+c_T(\beta_1 h - \beta_2 h\ln H )]}\, ,~~~~H = 1-h
\eea

A common feature to all of the foregoing functions is that the reduced area vanishes for $n \to \infty$. Physically this is not objectionable since the adsorption and hindered adsorption are negligible for $n>5$. But it may be useful to introduce a second parameter $\ga$ which gives a finite reduced area for $n \to \infty$ and thus makes it possible to control the initial changes from $\beta_0$ to $\beta_2$, etc. So we generalize Eq. (\ref{a21}); 
 \beq\label{a37}
  \beta_n = \gamma + \frac{1-\gamma}{ (1+n)^r }, ~~~~ n = 0,1,2,3,4,...
 \eeq
\bea \label{a38}
  && \sum_{n=1}^\infty \beta_n h^n = \sum_{n=1}^\infty \left[\gamma + \frac{1-\gamma}{ (1+n)^r } \right]h^n = \gamma\frac{h}{1-h} + (1-\gamma)\left(\frac{Li_r(h)}{h}-1\right)
 \\ \label{a39}  \mbox{and}~~
  && \sum_{n=1}^\infty \beta_n n h^n = \sum_{n=1}^\infty \left[\gamma + \frac{1-\gamma}{ (1+n)^r } \right] n h^n = \gamma\frac{h}{(1-h)^2} + (1-\gamma)\frac{Li_{r-1}(h)-Li_r(h)}{h}~~~~~~~~~
\eea
The resulting sorption isotherm is shown in Fig. \ref{fig3}d:
\bea \label{a40}
\tht(h,T)\ =\ c_T\,\frac{ \gamma h^2 +  (1-\gamma)H^2[Li_{r-1}(h)-Li_r(h)]}{hH^2+c_T H [ \gamma h^2 +  (1-\gamma)H(Li_{r}(h)-h)]}
\eea

However, not all analytical expressions can be obtained when $\beta_n$ is a general decreasing function of $n$.
Exponential functions, for example, are not suitable. For $\beta_n=e^{-\beta_1 n}$, the sum in the denominator of Eq. (\ref{a16}) converges only if $|h| < e^{-\beta_1}$, and the sum in the numerator cannot be expressed in terms of any known function. Besides, whereas numerical summation of the first few significant terms would be no problem, the exponential decay might be too fast for characterizing nanopore structures. Unlike the power functions, the exponentials are not self-similar, which means that some characteristic nanopore size would have to exist, but none can be identified.   

Another way to compare the foregoing functions $\tht(h,T)$ is the initial slope. This slope can be determined and compared against BET theory to show how fast the restriction on free surface area develops. If this slope deviates significantly from the BET theory, one can infer that a large amount of pore space has a width on the order of nanometer scale. However, if it is close to the initial slope of BET curve, then one can say that the first few layers (which are also dominant ones in the adsorption regime) are quite free to develop. Such initial behavior  can be obtained if we set the parameter $\beta_1,\beta_2$ in Eqs. (\ref{a28}),(\ref{a32}),(\ref{a36}) close to 1.

In general, $\tht(h,T)$ can be easily computed by evaluating, numerically, the sums in Eq. (\ref{a16}). Only the terms for $n = 0, 1, 2,...5$  need to be computed since the subsequent terms are negligible.

Unlike the BET isotherm, none of foregoing isotherms is amenable to linear regression. Nevertheless, the isotherm parameters can be identified from sorption data almost instantly, by using optimum fitting of the measured isotherm with a powerful nonlinear optimization algorithms such a the Levenberg-Marquardt and using the BET isotherm as the initial estimate. This isotherm is a special case for $r \to 0$.
Compared to \cite{brunauer1940theory}, the present isotherms are not only more generally applicable but also much simpler.

Brunauer {\em et al.} \cite{brunauer1940theory,brunauer1943adsorption} distinguished 5 types of isotherm shapes. The present isotherm formulae are of type II . However, with the present method, a transition between different sorption isotherm types are possible. For example, for $r \to infty$ in Eq. (\ref{a29}), one gets the classical Langmuir isotherm for monomolecular adsorption layers. When $c_T$ is very small, i.e., the adsorbent-solid interaction is very weak, the type III isotherm is retrieved. By proper calibration of $\beta_n$ and exponent  $r$ in Eq. (\ref{a21}), types IV and V can also be reproduced.

\subsection*{Qualitative Comparisons with Some Previously Observed or Simulated Isotherms }
To illustrate the behavior and usefulness of the present theory, the predicted isotherms and BET isotherms need to be compared with  experiments in which hindered adsorption is likely to happen. This can be either a normal Portland cement paste with low water-cement ratio or various special types of cement with extra minerals or crystals to be formed. In these adsorbents, the C-S-H platelets or needles grow and create nanopores in which full adsorption layers exposed to vapor cannot develop. Using the same procedure as in  \cite{brunauer1940theory}, the $c_T$, the monomolecular layer volume and mass, $v_m$ and $u_m$ (characterized as volume or mass adsorbed per 1 gram of dry sample), were obtained by fitting the regime of low relative humidity with BET theory. After that, the remaining relevant parameters (depending on which function was used) were fitted until the least-square error of fit was minimized.    

Wang {\em et al.}'s \cite{wang2012pore} studied a compacted pore structure of phosphoaluminate cement (PAC) paste with water-cement ratio $w/c = 0.32$. Under scanning electron microscopy (SEM) image, the formation of cotton-shaped gels was observed, resulting in relatively low porosity and small pore sizes. As a result, Fig. \ref{fig4}a displayed a significant deviation from the BET isotherm  above $h \approx 0.25$. A better match if obtained by fitting to these data the present isotherm with function $\beta_n$ given be Eq. (\ref{a21}) with parameters $\beta_1=0.998$ and $r = 0.986$. Obviously, the fit is better for $h>0.25$. 

In another experiment, Powers and Brownyard \cite{powers1946studies} measured adsorption for a type IV cement paste with $w/c = 0.309$ and the equivalent age of $t_e \approx 1$ year. Up to 0.5, BET gave a good fit, but overestimated the isotherm for $h>0.5$; 
see Fig. \ref{fig4}b.

\subsection*{Conclusions}
 
 \be \setlength{\itemsep}{-1.5mm} 
 \ii
Because the area of a multimolecular hindered adsorbed layer increases with the relative humidity of vapor, the area of the free (or unhindered) adsorption layers exposed to pore vapor decreases with the thickness of the multimolecular adsorbed layer, and thus also with the increasing humidity of vapor in the pores. 
 \ii
The basic idea in extending the BET isotherm to nano-porous solids with hindered adsorption is to treat the transitions from free to hindered adsorption as lateral constraints imposing an area reduction factor that decreases from one molecular adsorption layer to the next.
 \ii
The key point in the derivation of the isotherms is that the area reduction factors apply only to overall volume and area of the free adsorbed layers but not to the local rates of water evaporation, liquefaction and adsorption on the solid adsorbent surface.   
  \ii
Generally, the isotherm is obtained as the ratio of the sums of two infinite series.
 \ii
Considering the area reduction factor to be inversely proportional to the layer number leads to a simple analytical formula for the sorption isotherm. A general inverse power-law dependence of this factor on the layer number is also analytically tractable, and yields the isotherm expressed in terms of polylogarithm (aka Jonqui\`ere) functions. For the same initial slope, the resulting isotherms deviate from the BET isotherm downward. The deviation grows with increasing exponent.
 \ii
The inverse power law exponent is an additional empirical parameter providing flexibility in test data fitting. Excluding the fist layer from the power law can provide another fitting parameter.
  \ii 
Qualitative comparisons with some published isotherms observed experimentally on cement pastes 
indicate that the present theory modifies the BET isotherm in the right direction.
 \ii
The present analysis applies only to the drying or wetting that is not so slow as to allow significant mass exchange between water vapor and hindered adsorbed water in the nanopores less than 3 nm wide.
 \ee
 
\subsection*{Appendix. Comment on sorption slow enough for all nanopores to reach equilibrium}    
To consider this case, the sorption isotherm $\tht(h,T)$ may be redefined to represent only the volume of the free adsorbed layers directly exposed to vapor. The total relative volume of adsorbed water may be, in theory, approximated as
 \beq  \label{g1}
  \tht_{total}(h,T) = \tht(h,T) + 2 H(h - h_n)[1 - \beta_n(h)] 
 \eeq
where $H$ now represents the Heaviside step function and $h_n$ is the vapor humidity at which nanopores $2n$ molecules wide, in equilibrium, completely filled by $2n$ molecular layers of the adsorbate. This case, however, requires further research.  

\vv {\small \no {\bf Acknowledgment:}\, Partial financial support from the Department of Energy through Los Alamos National Laboratory grant number 47076 to Northwestern University is gratefully acknowledged. Preliminary research relevant for concrete was supported by the U.S. Department of Transportation through Grant 20778 from the Infrastructure Technology Institute of Northwestern University, and from the NSF under grant CMMI-1129449. 




\listoffigures  
\clearpage
\bfi
\centering
  \includegraphics[width=0.5\textwidth] {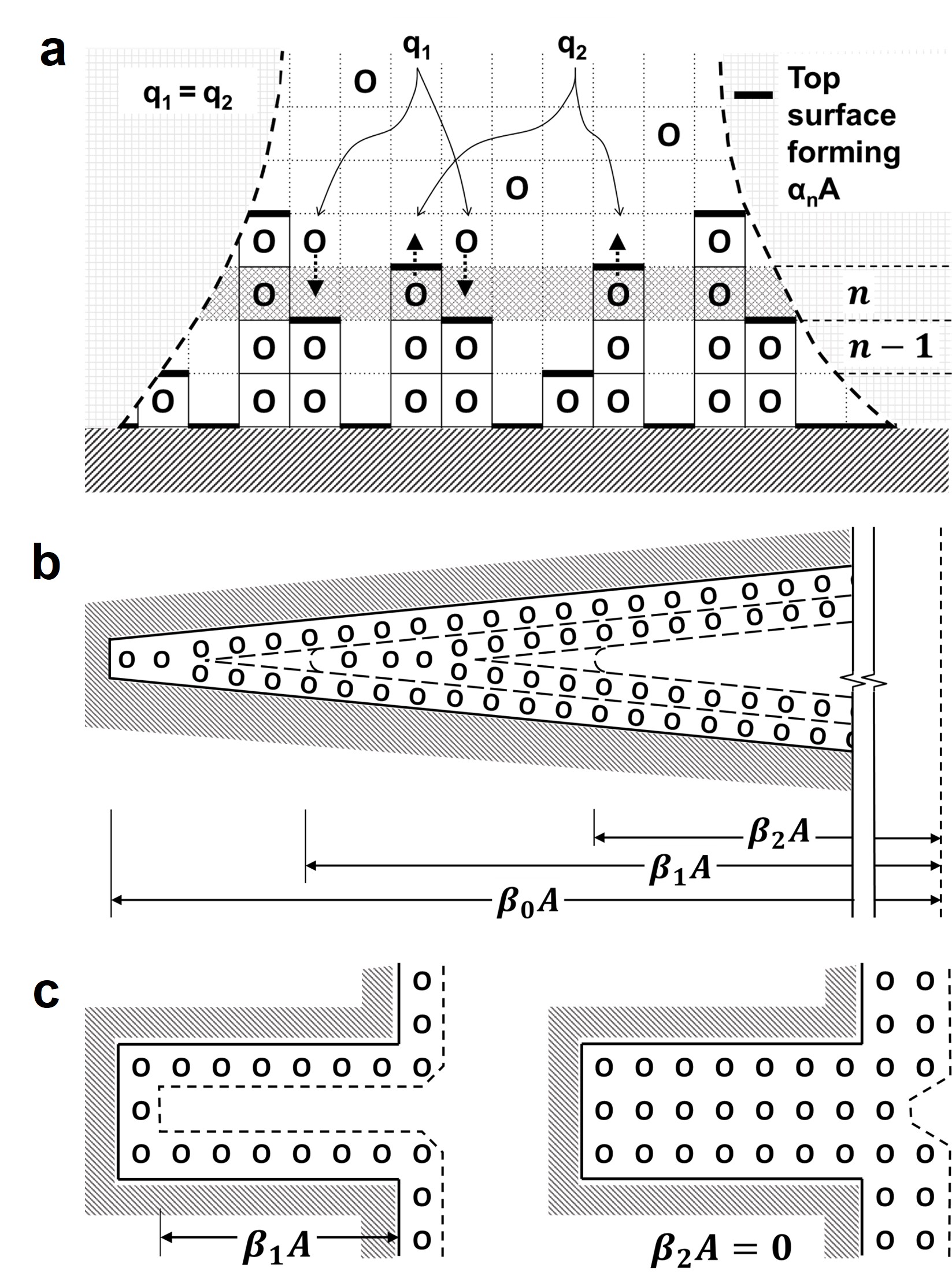}
 \caption{\label{fig2} \sf (a) Molecule movements during condensation and evaporation in the $n^{th}$ layer balancing each other; (b) Wedge nanopore containing the equivalent of one or two molecular adsorption layers, showing change of surface area exposed to vapor (unlike (a), randomness of molecular positions not shown here; (c) Varying of vapor exposed surface when the surface coverage of pores of uniform width changes from one to two molecules (we disregard the fact that the vapor would have to move along the nanopore under influence of surface forces).}
\efi
\bfi    
\centering       
  \includegraphics[width=\textwidth] {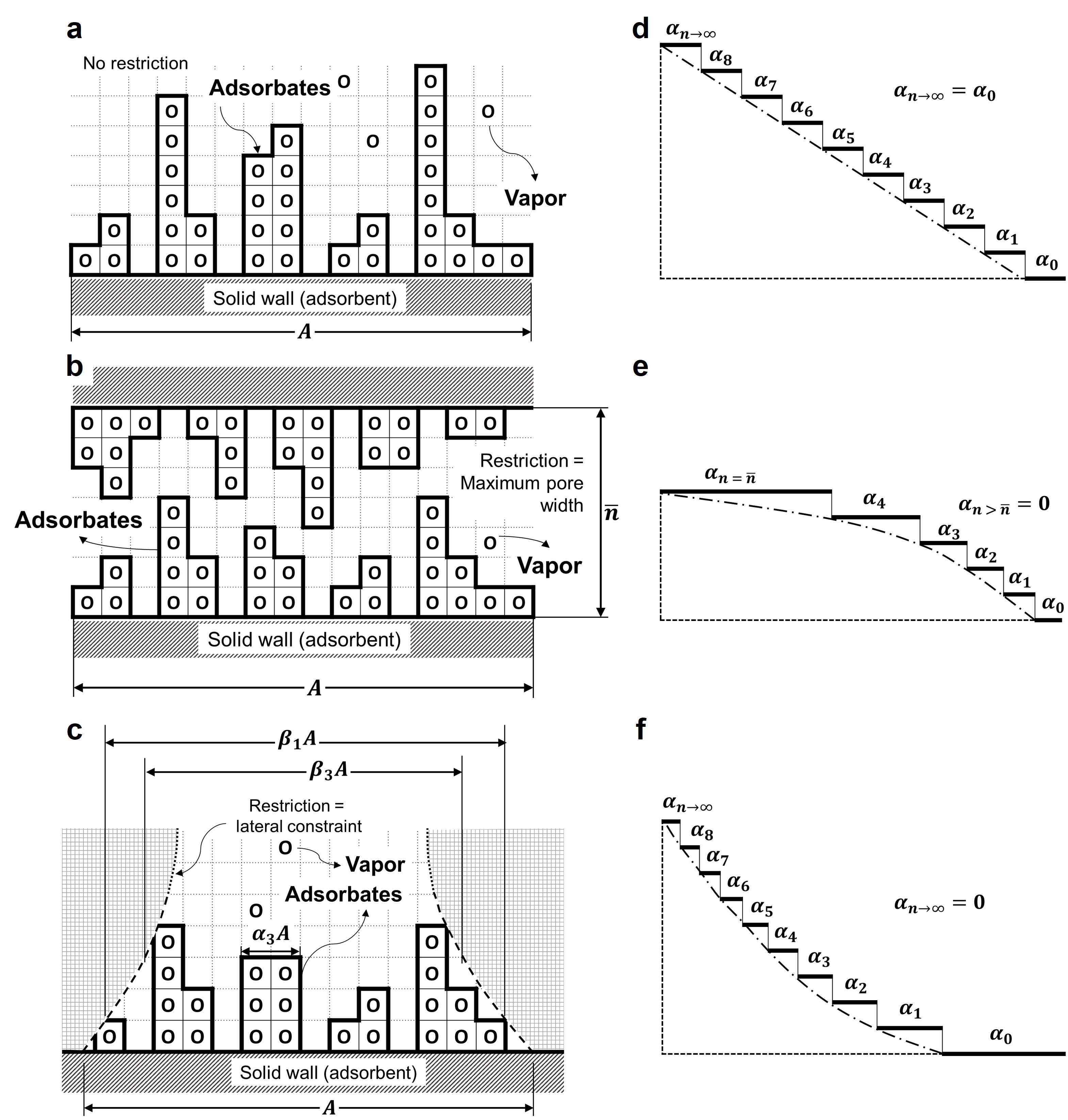}
\caption{\label{fig1} \sf The adsorption of vapor onto solid wall and corresponding shape of lateral constraint: (a) Example of random arrangement of adsorbed molecules BET theory where the vapor exposed surface area is equal to the adsorbent base area regardless of the number $n$ of layers, which is equivalent to the lateral constraint shown in (d); (b) interference of two-sided adsorption in nanopores with opposite parallel planar walls, considered in \cite{brunauer1940theory},
which is equivalent to the lateral constraint shown in (e); (c) array of adsorbed molecules with the lateral constraint of exposed surface due to hindered adsorption, and the corresponding shape of the lateral constraint shown in (e); figure (a) is adapted from \cite{brunauer1938adsorption}, and figure (b) from  \cite{brunauer1940theory}.} 
\efi
\bfi 
\centering
  \includegraphics[width=\textwidth] {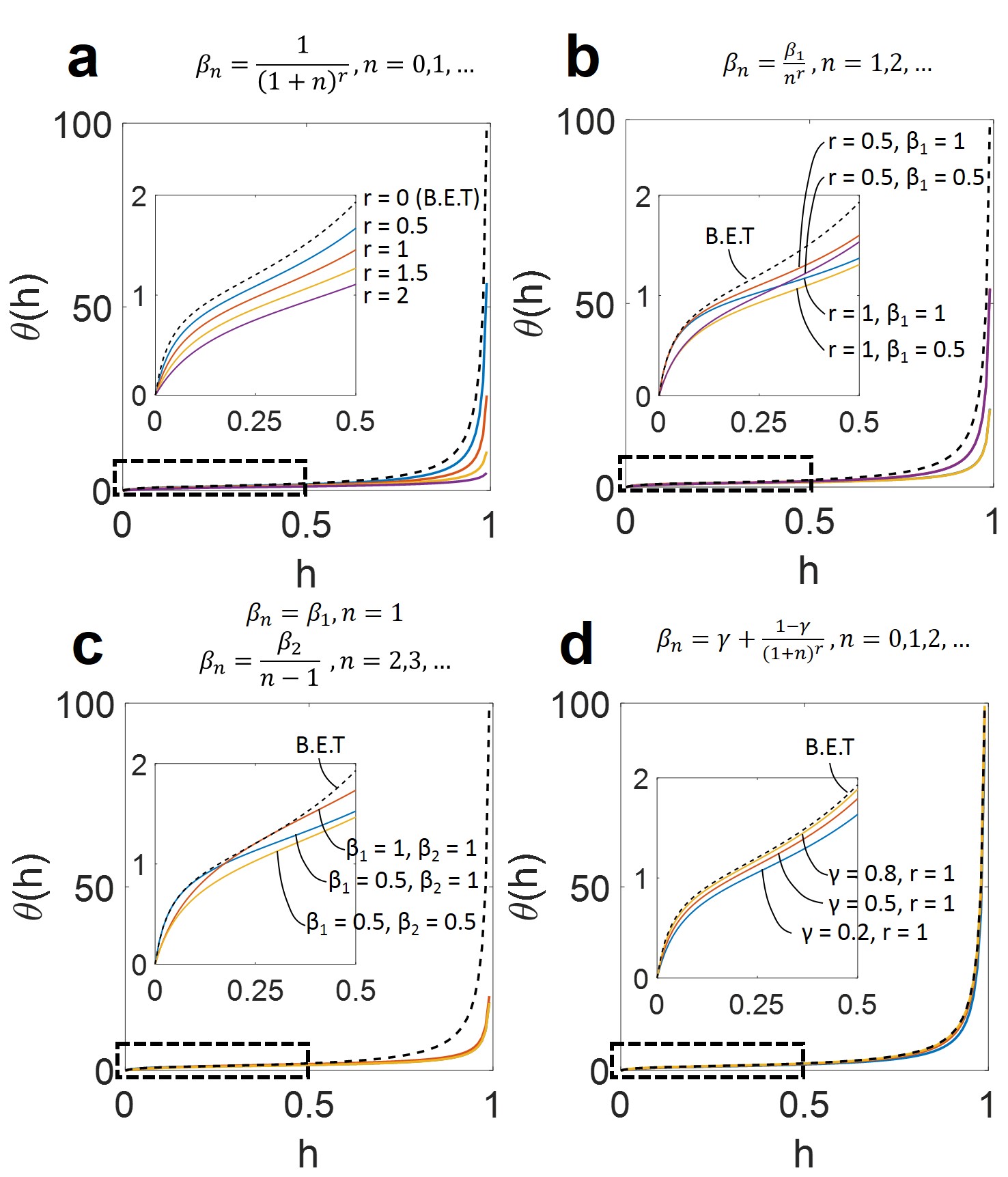}
 \caption{\label{fig3} \sf Adsorption Isotherms corresponding with Eqs. (\ref{a24}), (\ref{a32}), (\ref{a36}), and (\ref{a40}) ($c_T=28$) of the present hindered adsorption theory compared with BET prediction.}
\efi

\bfi 
\centering
  \includegraphics[width=\textwidth] {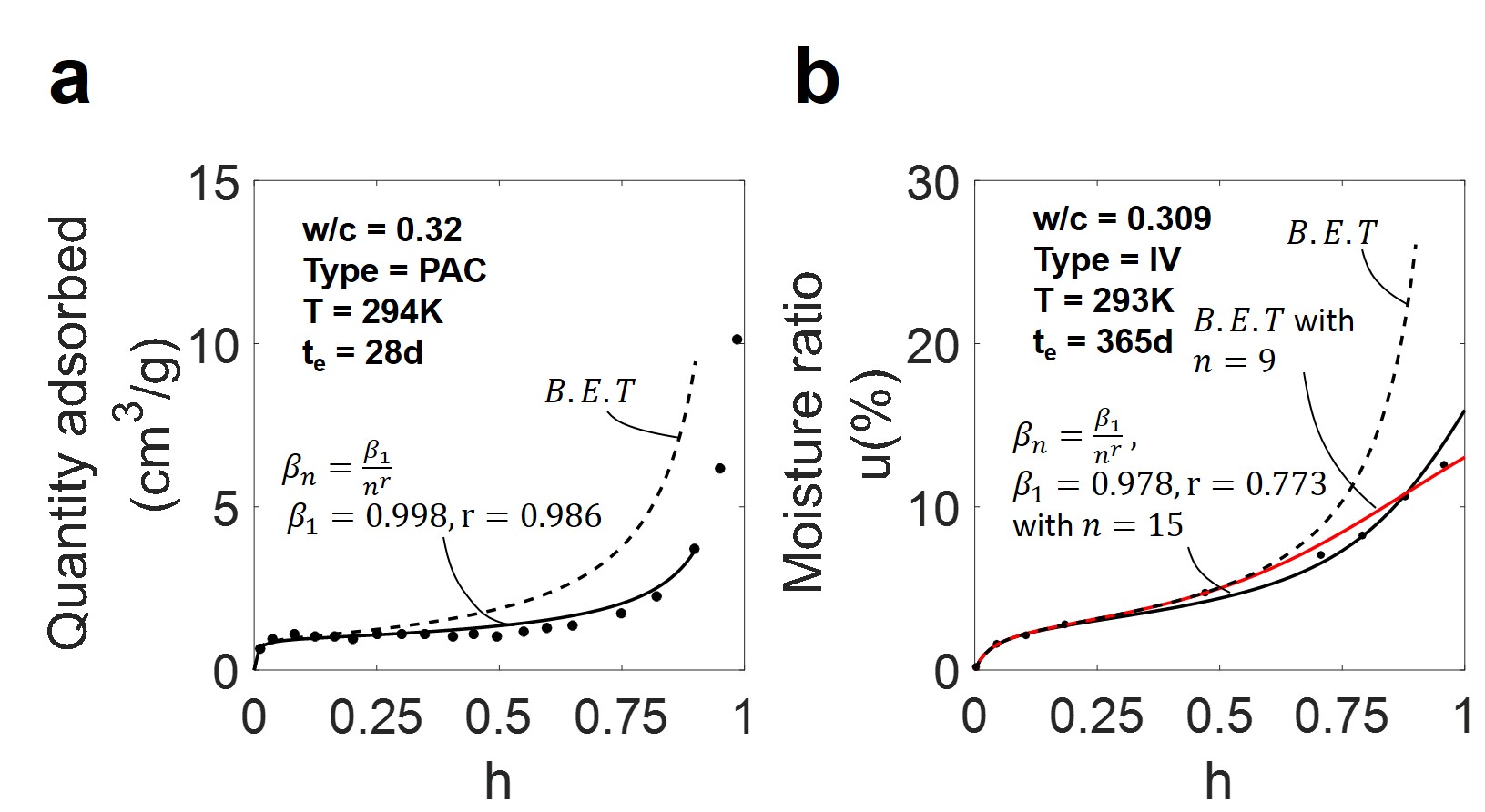}
 \caption{\label{fig4} \sf Fitting of published experimental results on cement pastes using the BET theory and the present theory; Tests: (a) Wang {\em et al.} \cite{wang2012pore} and (b) Powers and Brownyard \cite{powers1946studies}.}
\efi

\vskip 4mm

\bibliographystyle{unsrt}

\end{document}